\documentclass[prl,aps,showpacs,twocolumn,superscriptaddress,amsmath,amssymb]{revtex4}

\usepackage[dvips]{graphicx}
\DeclareGraphicsExtensions{.ps}

\newcommand*{\etal}{\textit{et al.}}

\begin{document}

\title{Quantum imaging with incoherent photons}
\date{\today}

\author{C. Thiel}
\affiliation{Institut f\"ur Optik, Information und Photonik, Universit\"at Erlangen-N\"urnberg, 91058 Erlangen, Germany}

\author{T. Bastin}
\affiliation{Institut de Physique Nucl\'eaire, Atomique et de Spectroscopie, Universit\'e de Li\`ege au Sart Tilman, 4000 Li\`ege, Belgium}

\author{J. Martin}
\affiliation{Institut de Physique Nucl\'eaire, Atomique et de Spectroscopie, Universit\'e de Li\`ege au Sart Tilman, 4000 Li\`ege, Belgium}

\author{E. Solano}
\affiliation{Physics Department, ASC, and CeNS, Ludwig-Maximilians-Universit¨at, Theresienstrasse 37, 80333 Munich, Germany}
\affiliation{Secci\'on F\'isica, Departamento de Ciencias, Pontificia Universidad Cat\'olica del Per\'u, Apartado Postal 1761, Lima, Peru}

\author{J. von Zanthier}
\affiliation{Institut f\"ur Optik, Information und Photonik, Universit\"at Erlangen-N\"urnberg, 91058 Erlangen, Germany}

\author{G. S. Agarwal}
\affiliation{Department of Physics, Oklahoma State University, Stillwater, OK 74078-3072, USA}



\begin{abstract}
We propose a technique to obtain sub-wavelength resolution in
quantum imaging with potentially $100 \%$ contrast using incoherent light. Our
method requires neither path-entangled number states nor
multi-photon absorption. The scheme makes use of $N$ photons
spontaneously emitted by $N$ atoms and registered by $N$ detectors.
It is shown that for coincident detection at particular detector
positions a resolution of $\lambda / N$ can be achieved.
\end{abstract}

\maketitle

In Young's double slit experiment (or in a Mach-Zehnder
interferometer) the probability $G^{(1)}({\mathbf r})$ to detect a
photon at position ${\mathbf r}$ results from the interference of
the two possible paths a single photon can take to reach the
detector. This is expressed by the state $| \psi (1) \rangle =
1/\sqrt{2} \, (|1\rangle_U |0\rangle_L + |0\rangle_U |1\rangle_L)$
where the subscript $L$ ($U$) denotes the lower (upper) arm of the
interferometer. Variation of the detector position leads to a
modulation of the form $G^{(1)}({\mathbf r}) \propto 1 + \cos\delta
({\mathbf r})$, where $\delta ({\mathbf r}) = k d \sin \theta
({\mathbf r})$ is the optical phase difference of the waves
emanating from the two slits and $k$, $d$ and $\theta ({\mathbf r})$
are the wavenumber, slit separation and scattering angle,
respectively. Obviously, the fringe spacing of the modulation (in
units of $d \sin \theta ({\mathbf r})$) is determined by the optical
wavelength $\lambda$, in correspondence with the Rayleigh criterion~\cite{Born}.

Quantum entanglement is able to bypass the Rayleigh limit
\cite{Fonseca99,Dowling00,Shih01,Itoh02,Boyd04,Zeilinger04,Scully04,Bollinger96,Agarwal03,
footnote1}. Consider for example the path-entangled $N$-photon state
$| \psi (N) \rangle = 1 / \sqrt{2} \, (|N\rangle_U |0\rangle_L +
|0\rangle_U |N\rangle_L)$. Because the $N$-photon state $|N\rangle$ has $N$ times the energy of
the single photon state $|1\rangle$ in a given mode it accumulates
phase $N$ times as fast when propagating through the setup. This
gives rise to an $N$ photon absorption rate of the form
$G^{(N)}({\mathbf r}, \ldots, {\mathbf r}) \propto 1 + \cos N \delta
({\mathbf r})$ exhibiting a fringe spacing $N$ times narrower than
that of $G^{(1)}({\mathbf r})$ \cite{Dowling00}. This gain in resolution can be fruitfully applied for a wide range of applications, e.g., in lithography~\cite{Dowling00,Shih01}, microscopy~\cite{Scully04}, spectroscopy \cite{Bollinger96} and even magnetometry~\cite{Agarwal03}. In order to implement this $N$-fold increase in resolution commonly an entangled state of the form $| \psi(N) \rangle$ in combination with a non-linear medium sensitive to $N$-photon absorption is  needed \cite{footnote1}. 

In this letter we propose a different scheme to achieve a resolution
of $\lambda / N$ involving neither of the above requirements. In
what follows we will apply this scheme in the context of microscopy.
The method employs $N$ photons spontaneously emitted from $N$ atoms
subsequently detected by $N$ detectors where by means of 
post-selection it is ensured that precisely one photon is recorded at each of
the $N$ detectors. We demonstrate that in this case, for certain detector
positions ${\mathbf r}_2, \ldots, {\mathbf r}_N$, the $N$th order
correlation function as a  function of ${\mathbf r}_1$ takes
the form $1 + \cos N \delta ({\mathbf r}_1)$, resulting in a phase
modulation with a theoretical contrast of
$100 \%$ and a fringe spacing determined by $\lambda / N$.
As with path-entangled number states, this corresponds to
an $N$-fold reduced fringe spacing compared to $G^{(1)}({\mathbf
r})$ while keeping a contrast of potentially $100 \%$. 
Hereby, only tools of linear optics are employed as a single photon is registered at each detector.

To understand this outcome let us consider $N$
identical two-level atoms excited by a single laser $\pi$-pulse. After the spontaneous emission the $N$ photons are registered by $N$ detectors at positions ${\mathbf r}_1, \ldots {\mathbf r}_N$. For the sake of simplicity let us consider coincident detection~\cite{comment1}. In that case the $N$th order correlation function~\cite{Glauber63} can be written (up to an insignificant prefactor) as~\cite{Agarwal74}
\begin{equation} \label{G(N)}
G^{(N)}({\mathbf r}_1, \ldots , {\mathbf r}_N) = \langle D^{\dagger}
({\mathbf r}_1) \ldots D^{\dagger} ({\mathbf r}_N) D({\mathbf r}_N)
\ldots D({\mathbf r}_1) \rangle
\end{equation}
where
\begin{equation} \label{detector}
D({\mathbf r}_i) = \frac{1}{\sqrt{N}} \sum_{\alpha = 1 }^{N}
\sigma^{-}_{\alpha} e^{-i k {\mathbf n}({\mathbf r}_i) \cdot {\mathbf
R}_{\alpha}} .
\end{equation}
Here ${\mathbf n}({\mathbf r}_i) = {\mathbf r}_i / r_i$ stands for the unit vector in the direction of
detector $i$, the sum is over all atom positions ${\mathbf R}_{\alpha}$, $k = \omega_0 /c$, with $\omega_0$ the transition frequency, and $\sigma^{-}_{\alpha} =
|g\rangle_{\alpha} \langle e |$ is the lowering operator of atom $\alpha$ for the
transition $| e\rangle\rightarrow {| g\rangle}$.

For all atoms initially prepared in the excited state $|e\rangle$,
we obtain from Eqs.~(\ref{G(N)}) and~(\ref{detector})
\begin{equation} \label{G(N)_new1}
G^{(N)}({\mathbf r}_1,...,{\mathbf r}_N)= \frac{1}{N^{N}} |\gamma({\mathbf r}_1,...,{\mathbf r}_N)|^2,
\end{equation}
where~\cite{Agarwal04}
\begin{equation} \label{gamma}
\renewcommand{\arraystretch}{0.8}
\gamma({\bf r}_1,..., {\bf r}_N)= \sum\limits_{\begin{array}{c}
\textrm{\scriptsize $\epsilon_1, \ldots , \epsilon_N = 1$} \\
\textrm{\scriptsize $\epsilon_1 \neq \ldots \neq \epsilon_N$}
\end{array}}^N \prod\limits^N_{\alpha=1}e^{- ik\,{\bf n}({\mathbf
r}_{{\mbox{\scriptsize$\epsilon$}}_{\alpha}})\cdot{\bf R}_{\alpha}} \, .
\renewcommand{\arraystretch}{1}
\end{equation}
Equations~(\ref{G(N)_new1}) and~(\ref{gamma}) show that $G^{(N)}({\mathbf
r}_1,...,{\mathbf r}_N)$ results from the interference of $N!$
terms, associated with all possibilities to scatter $N$
photons from $N$ identical atoms which are subsequently registered by $N$ detectors.

\begin{figure}
\begin{center}
\noindent\mbox{\includegraphics[width=6cm, bb=43 96 416 329,
clip=true]{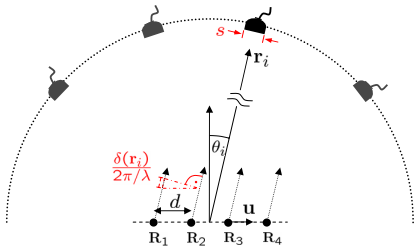}}
\end{center}
\vspace{-0.6cm} \caption{(Color online) Atomic arrangement and detection scheme:
$N$ identical two-level atoms at ${\mathbf R}_1,  \ldots, {\mathbf
R}_N$ spontaneously emit $N$ photons after excitation by a laser
pulse. The photons are recorded in the far field by $N$ detectors
positioned at ${\mathbf r}_1,  \ldots, {\mathbf r}_N$. The figure
exemplifies the case $N = 4$ (for additional symbols see text).}
\label{Fig1}
\end{figure}

To simplify further calculations let us consider the case of
$N$ equidistant atoms.
By choosing the origin of the coordinate system in the center of the atomic chain, we have
\begin{equation} \label{R}
{\mathbf R}_{\alpha} = j_{\alpha} d {\mathbf u}
\end{equation}
with ${\mathbf u}$ the unit vector along the chain axis, $d$ the interatomic spacing and
$j_{\alpha}= $ $-(N-1)/2, \ldots , (N-1)/2$ for $\alpha = 1, \ldots, N$ (see Fig.~1). By defining
\begin{equation} \label{delta}
\delta ({\mathbf r}_i) = k d {\mathbf n}({\mathbf r}_i) \cdot {\mathbf u} = kd\sin \theta_i
\end{equation}
where $\theta_i$ is the angle between ${\mathbf n} ({\mathbf r}_i)$ and
the direction normal to the atomic chain (see Fig.~1), we find
\begin{equation} \label{G(N)_1}
G^{(N)}({\mathbf r}_1, \ldots ,{\mathbf r}_N)= \frac{1}{N^N} \left[
\sum \cos
({\mathbf j} \cdot  {\boldsymbol \delta}) \right]^2 \, .
\end{equation}
Here, ${\mathbf j}$ is the vector of the distances of the atoms from the origin in units of $d$:
\begin{equation} \label{j}
{\mathbf j} =  (j_1, \ldots , j_N) \, ,
\end{equation}
${\boldsymbol \delta}$ is given by:
\begin{equation} \label{delta1}
{\boldsymbol \delta} =  ( \delta({\mathbf r}_1), \ldots , \delta({\mathbf r}_N)) \,,
\end{equation}
and the sum in Eq.~(\ref{G(N)_1}) is over the $N!$ permutations of the $\mathbf j$ components.

Due to the symmetry of the configuration, the function $G^{(N)}({\mathbf r}_1, \ldots ,{\mathbf r}_N)$ contains $N!/2$ cosine terms, each oscillating in general with a different spatial frequency. Obviously, the complexity of the
expression rises rapidly with increasing atom number $N$. However, if the $N$ detectors are placed in such a manner that all terms in Eq.~(\ref{G(N)_1}) interfere to give a single cosine, one is left with a modulation oscillating at a {\it unique} spatial frequency. This occurs in the following case:
for arbitrary {\it even} $N$ and choosing the detector positions such that
\begin{eqnarray} \label{detectoreven}
&& \delta({\mathbf r}_2) = - \delta({\mathbf r}_1) \, , \nonumber\\
&& \delta({\mathbf r}_{3}) =  \delta({\mathbf r}_{5}) = \ldots =  \delta({\mathbf r}_{N-1})=\frac{2 \pi}{N} \, , \, \nonumber\\
&& \delta({\mathbf r}_{4}) =  \delta({\mathbf r}_{6}) = \ldots =  \delta({\mathbf r}_{N}) = -\frac{2 \pi}{N} \, , \,
\end{eqnarray}
the $N$th order correlation function $G^{(N)}$ as a function of
detector position ${\mathbf r}_1$ reduces to
\begin{equation} \label{Gneven}
G^{(N)}({\mathbf r}_1) = A_N \, [ 1+\cos(N \, \delta({\mathbf r}_1))],
\end{equation}
where $A_N$ is a constant which depends on $N$. For arbitrary {\it
odd} $N>1$, and choosing the detector positions such that
\begin{eqnarray} \label{detectorodd}
&& \delta({\mathbf r}_2) = - \delta({\mathbf r}_1) \, , \nonumber\\
&& \delta({\mathbf r}_{3}) =  \delta({\mathbf r}_{5}) = \ldots =  \delta({\mathbf r}_{N})=\frac{2 \pi}{N+1} \, , \, \nonumber\\
&& \delta({\mathbf r}_{4}) =  \delta({\mathbf r}_{6}) = \ldots =  \delta({\mathbf r}_{N-1}) = -\frac{2 \pi}{N+1} \, , \,\end{eqnarray}
the $N$th order correlation function $G^{(N)}$ as a function of
${\mathbf r}_1$ reduces to
\begin{equation} \label{Gnodd}
G^{(N)}({\mathbf r}_1) = A_N \, [ 1+\cos((N+1) \, \delta({\mathbf r}_1))].
\end{equation}

According to Eqs.~(\ref{Gneven}) and~(\ref{Gnodd}), we are able to obtain
for any $N$ a correlation signal with a modulation of a single
cosine, displaying a contrast of $100 \%$ and a fringe spacing determined 
by $\lambda/N$ ($\lambda/(N+1)$) for even (odd) $N$. 
We note that due to the limited detector sizes and the dipole emission pattern of the
spontaneously emitted photons only a subset of all emitted photons
will be recorded. However, in contrast to using maximally path-entangled $N$-photon states we
are able to avoid in this scheme both the necessity to generate a state of the form $| \psi (N) \rangle$ and the need to detect a multi-photon absorption signal~\cite{comment2}. We emphasize that as the photons are produced by spontaneous decay the interference signal is generated by incoherent light. We stress further that a fringe contrast implied by Eq.~(\ref{Gneven}) or Eq.~(\ref{Gnodd}) proves the underlying quantum nature of the process~\cite{Mandel83,Ou88,Skornia01}. The quantum character is generated by the measurement process after the detection of  the first photon. In fact, just before the detection of the $N$th photon, the atomic system is in an $N$-particle $W$-state with one excitation~\cite{Wstate}. The non-classical characteristics of our scheme are thus another example of detection induced entanglement of initially uncorrelated distant particles~\cite{Cabrillo99,Plenio99,Polzik99,Skornia01,Kimble05,Grangier06,Monroe06}.

To exemplify our method, let us consider the simplest situation, i.e., the case of $N = 2$ atoms. With ${\mathbf j} = (-\frac{1}{2}, + \frac{1}{2})$
we obtain from Eq.~(\ref{G(N)_1})
\begin{equation} \label{G2twoions}
G^{(2)}({\mathbf r}_1, {\mathbf r}_2) = \frac{1}{2} \left[ 1 +  \cos (\delta({\mathbf r}_1) -
\delta({\mathbf r}_2)) \right].
\end{equation}
Obviously, the modulation of the $G^{(2)}({\mathbf r}_1,
{\mathbf r}_2)$-function depends on the relative position of the two
detectors (see Fig.~2): for $\delta({\mathbf r}_2) = \delta({\mathbf r}_1)$ the
second order correlation function is a constant, whereas for fixed
$\delta({\mathbf r}_2)$ the two photon coincidence as a function of
$\delta({\mathbf r}_1)$ exhibits the same phase modulation and
fringe spacing as $G^{(1)}({\mathbf r})$ in Young's double slit experiment. However, the increased
parameter space available for the detector positions in case of two
detectors allows also to pick out the relative orientation
$\delta({\mathbf r}_2) = - \delta({\mathbf r}_1)$. In this case we
get
\begin{equation} \label{G2twoions1}
G^{(2)}({\mathbf r}_1) = \frac{1}{2} \left[ 1 +  \cos
(2 \delta({\mathbf r}_1)) \right] ,
\end{equation}
exhibiting a phase modulation as a function of ${\mathbf r}_1$ with {\it half} the fringe spacing of
$G^{(1)}({\mathbf r})$ while keeping a contrast of 100$\%$ (see also~\cite{Agarwal04}). Note that the assumed condition for the direction of emission of the
two photons, i.e., $\delta({\mathbf r}_2) = - \delta({\mathbf r}_1)$,
corresponds to a space-momentum correlation of the photons identical
to the one present in spontaneous parametric down conversion
\cite{Shih01,Shih03,Boyd04}.

\begin{figure}
\begin{center}
\noindent\mbox{\includegraphics[width=8.5cm, bb=28 229 563 805,
clip=true]{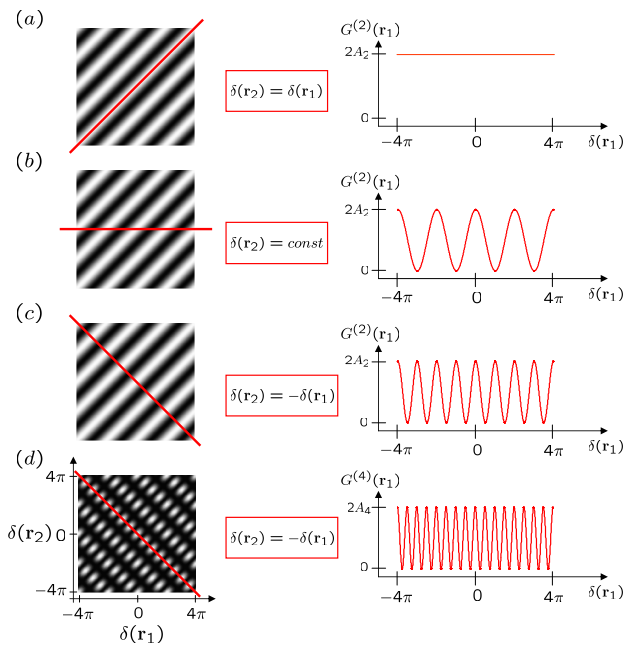}}
\end{center}
\vspace{-0.6cm} \caption{(Color online) Left (a) - (c): density plots of $G^{(2)}({\mathbf r}_1, {\mathbf r}_2)$ for two atoms versus $\delta({\mathbf r}_1)$ and $\delta({\mathbf r}_2)$; left (d): density plot of $G^{(4)}({\mathbf r}_1, {\mathbf r}_2, {\mathbf r}_3, {\mathbf r}_4)$ for four atoms versus $\delta({\mathbf r}_1)$ and $\delta({\mathbf r}_2)$, with $\delta({\mathbf r}_3) = \pi/2$ and $\delta({\mathbf r}_4) = -\pi/2$. Right: cuts through the density plots along the indicated lines, i.e., for (a) $\delta({\mathbf r}_2) = \delta({\mathbf r}_1)$, (b) $\delta({\mathbf r}_2) = const$ and (c), (d) $\delta({\mathbf r}_2) = - \delta({\mathbf r}_1)$.} \label{Fig2}
\end{figure}

In the case of the fourth order correlation function
$G^{(4)}({\mathbf r}_1, {\mathbf r}_2, {\mathbf r}_3, {\mathbf
r}_4)$ for four equidistant atoms, and by placing the detectors 
according to Eq.~(\ref{detectoreven}) (see Fig.~2), one finds
\begin{equation} \label{G(44)single}
G^{(4)}({\mathbf r}_1) = \frac{1}{8} \, \left[ 1+\cos ( 4 \, \delta({\mathbf r}_1) ) \right] \, .
\end{equation}
Obviously, $G^{(4)}$ as a function of ${\mathbf r}_1$ exhibits a
modulation of a single cosine with a contrast of 100$\%$, in this
case with a fringe spacing determined by $\lambda/4$.

\begin{figure}
\begin{center}
\noindent\mbox{\includegraphics[width=6cm, bb=180 350 435 600,
clip=true]{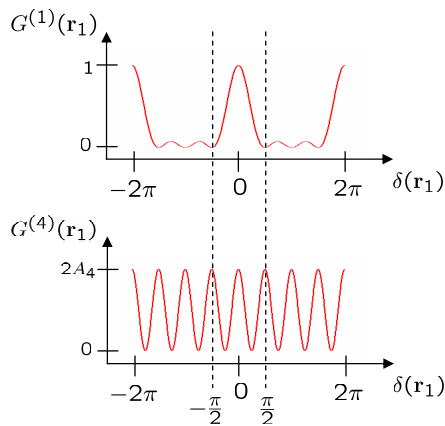}}
\end{center}
\vspace{-0.6cm} \caption{(Color online) $G^{(1)}({\mathbf r}_1)$ and
$G^{(4)}({\mathbf r}_1)$ as a function of $\delta({\mathbf r}_1)$
for a chain of 4 atoms. The interval $[-2\pi,2\pi]$
corresponds to the maximal range of variation of $\delta({\mathbf
r}_1)$ for an interatomic distance $d = \lambda$ [see
Eq.~(\ref{delta})]. The dashed lines indicate the corresponding range 
in case of $d = \lambda/4$.} \label{Fig3}
\end{figure}

As an example, let us apply our scheme in the context of microscopy. From Abbe's 
theory of the microscope we know that an object can be resolved only 
if at least two principal maxima of the diffraction pattern are included 
in the image formation~\cite{Born}. 
According to this criterion the use of the first order correlation 
function $G^{(1)}({\mathbf r}_1)$ for imaging $N$ equidistant atoms 
allows at best to resolve an interatomic spacing equal to
$\lambda$~\cite{Born}. Indeed, if each atom is initially prepared in
the state $| \phi \rangle = \frac{1}{\sqrt{2}}(|g\rangle +
|e\rangle)$, we get from Eqs.~(\ref{G(N)}) and~(\ref{detector})
\begin{equation} \label{G1Nions}
G^{(1)}({\mathbf r}_1) = \frac{1}{2} \left[ 1 + \frac{1}{N}
\sum_{\alpha=1}^{N-1} (N-\alpha) \cos ( \alpha \delta({\mathbf r_1})
) \right] \, .
\end{equation}
Equation~(\ref{G1Nions}) equals (up to an offset) the outcome of the
classical grating. As is well-known from the grating equation 
(and as Eq.~(\ref{G1Nions}) explicitly shows) two principal 
maxima appear in the far-field diffraction pattern only if
the interatomic distance is greater or equal to $\lambda$ (see
Fig.~\ref{Fig3}). By contrast, the use of the $N$th order
correlation function with the $N$ detectors positioned according to
Eq.~(\ref{detectoreven}) (or Eq.~(\ref{detectorodd})) allows to
resolve an atom-atom separation as small as $\lambda/N$ 
(or $\lambda/(N+1)$) (see Fig.~\ref{Fig3}). In this way the $N$th 
order correlation function $G^{(N)}({\bf r}_1)$ 
can be used to resolve and image trapped atoms separated by a distance $d = \lambda/N$.

Finally, we address the technical feasibility of our scheme. For the 
ability to localize atoms and adequately resolve optical path differences 
on a scale smaller than $\lambda$ we refer 
to~\cite{Wineland93,Rauschenbeutel06,Monroe06a}. 
A detector of a given size $s$ 
positioned at a distance $L=|{\bf r}_i|$ in the far-field region 
(see Fig.~\ref{Fig1}) gives rise to an angular resolution 
$\Delta\theta=s/L$, i.e.\ to a phase resolution $\Delta \delta = kd \cos \theta \Delta \theta$. 
To resolve the modulation of the $N$-th order correlation 
function $G^{(N)}({\mathbf r}_1)$, a sufficient requirement is that 
$N \Delta \delta \ll 2 \pi$, i.e.\ $\Delta \theta \ll \lambda/(N d)$, 
which yields the condition
\begin{equation}
\label{Lcondition}
    L \gg s \frac{N d}{\lambda}.
\end{equation}
For given $N$ and $d$ we can thus find for any detector size $s$ a distance $L$ 
to achieve the necessary resolution. Hereby, choosing the smallest $L$ compatible with Eq.~(\ref{Lcondition}) 
is favorable in order to maximize the $N$ photon detection probability; the exact longitudinal positions of the detectors are thereby not important. In case of a gaussian distribution of the phases $\delta({\mathbf r_i})$ ($i = 2,\ldots, N$) around their ideal values given by Eqs.~(\ref{detectoreven}) or (\ref{detectorodd}) with a standard deviation $\sigma$ the contrast of the $G^{(N)}({\mathbf r}_1)$-function is reduced and given by $e^{-N \,\sigma^2/4}$. For $N=2$ and $N=4$, this means that a contrast of higher than 50\% can be maintained as long as $\sigma$ is less than 0.8 and 1.2, respectively. Using the set of reasonable parameters $d=5\, \mu$m, $\Delta d=0.1 \, \mu$m, $\theta({\bf r}_1)=30^{\circ}$, $\Delta\theta({\bf r}_1)=0.1^{\circ}$, $k=2\pi/800 \,$nm,
 $\Delta k < 10^{-7} \, k$ we obtain $\sigma \approx 0.7$.

In conclusion we have shown that $N$ photons of wavelength $\lambda$ spontaneously emitted by $N$ atoms and coincidentally recorded by $N$ detectors at particular positions exhibit correlations and interference properties similar to classical coherent light of wavelength $\lambda/N$. The method requires neither initially entangled states nor multi-photon absorption, only common single-photon detectors.

We gratefully acknowledge financial support by the Dr.~Hertha und
Helmut Schmauser foundation. G.S.A. thanks NSF grant no NSF-CCF-0524673 for supporting this collaboration. E.S. acknowledges support from DFG SFB 631, EU EuroSQIP projects, and the German
Excellence Initiative via the ``Nanosystems Initiative Munich''.


\begin{thebibliography}{99}
\bibitem{comment1}
We emphasize that coincident detection is not a prerequisite for our scheme. The requirement is that all photons emitted by the $N$ atoms are recorded by the $N$ detectors; the exact detection time does not influence the contrast nor the resolution of the correlation signal.
\bibitem{comment2}
Note that conditions~(\ref{detectoreven}) and~(\ref{detectorodd}) do not mean that the different detectors have to be at the same place since equality of phase shifts does not mean equality of detector positions.
\end{thebibliography}
\end{document}